\documentclass[pra,superscriptaddress,twocolumn,english, longbibliography, showpacs=false, nofootinbib]{revtex4-2}

\usepackage[T1]{fontenc}
\usepackage[utf8]{inputenc}
\usepackage{color}
\usepackage{babel}
\usepackage{amsmath}
\usepackage{amssymb}
\usepackage{amsthm}
\usepackage{subfigure}
\usepackage{graphicx}
\usepackage{physics} 
\usepackage{dsfont}
\usepackage{hyperref}
\usepackage{aurical}
\usepackage{float}

\DeclareMathAlphabet{\mathcal}{OMS}{cmsy}{m}{n}

\begin{document}

\title{Landauer principle and the second law in a relativistic communication scenario}

\author{Yuri J. Alvim}
\affiliation{QPequi Group, Institute of Physics, Federal University of Goi\'as, Goi\^ania, Goi\'as, 74.690-900, Brazil}

\author{Lucas C. C\'eleri\href{https://orcid.org/0000-0001-5120-8176}{\includegraphics[scale=0.05]{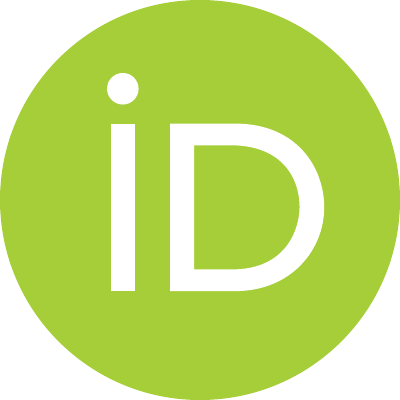}}}
\email{lucas@qpequi.com}
\affiliation{QPequi Group, Institute of Physics, Federal University of Goi\'as, Goi\^ania, Goi\'as, 74.690-900, Brazil}

\begin{abstract}
The problem of formulating thermodynamics in a relativistic scenario remains unresolved, although many proposals exist in the literature. The challenge arises due to the intrinsic dynamic structure of spacetime as established by the general theory of relativity. With the discovery of the physical nature of information, which underpins Landauer's principle, we believe that information theory should play a role in understanding this problem. In this work, we contribute to this endeavour by considering a relativistic communication task between two partners, Alice and Bob, in a general Lorentzian spacetime. We then assume that the receiver, Bob, reversibly operates a local heat engine powered by information, and seek to determine the maximum amount of work he can extract from this device. Since Bob cannot extract work for free, by applying both Landauer's principle and the second law of thermodynamics, we establish a bound on the energy Bob must spend to acquire the information in the first place. This bound is a function of the spacetime metric and the properties of the communication channel.  
\end{abstract}

\maketitle
 
\section{Introduction} 

The search for a relativistic theory of thermodynamics has a long history dating back to Einstein~\cite{Einstein} and Planck~\cite{Planck}, shortly after the discovery of special relativity. Despite ongoing debates and controversies~\cite{Ott,Landsberg}, the last century has seen numerous advancements exploring the connections between thermodynamics and both special and general relativity (see, e.g., Refs.~\cite{Tolman,Cavalleri,Newburgh,Landsberg2004,Dunkel,Jacobson,Eling,Rovelli1993,Rovelli2011,Rovelli2013,Padmanabhan,Basso}). However, achieving consensus remains elusive, particularly when considering the implications of general relativity.

Quantum information theory, due to its profound connection with thermodynamics~\cite{Wilde}, offers promising insights into this long-standing problem. Since Landauer's recognition that information is physical and subject to the laws of physics, information theory has become inherently linked to all branches of physics~\cite{Landauer}. The foundational Landauer principle asserts that erasing information necessitates dissipating energy as heat, a principle fundamental to both classical and quantum mechanics, and crucial to resolving Maxwell's demon paradox~\cite{Bennett,Benett2003}. Subsequent theoretical~\cite{Georgescu2021,Xu2024,Zivieri2022,Reeb2004,Norton2005,Norton2011,Esposito2011,Goold2015,Hilt2011,
Ladyman2007,Barnett2013,Maroney2009,Timpanaro2020} and experimental~\cite{Berut2012,Jun2014,Peterson2016,Hong2016} developments have followed these seminal works.

In this study, we explore the intersection of these ideas within the context of relativistic communication. By applying Landauer's principle alongside the second law of thermodynamics, we establish bounds on the energy involved in communication tasks. Specifically, we investigate the transmission of classical information between two parties using a massless scalar quantum field in a general curved spacetime, as discussed in Refs.~\cite{Landulfo2016,Landulfo2021}. Building upon this framework, we consider scenarios where one party, Bob, operates a heat engine powered by the information received from the communication channel. This setup enables Bob to locally extract work using the acquired information.

We demonstrate that since only information, not energy, is transmitted via the channel~\cite{Landulfo2021}, both Landauer's principle and the second law imply that Bob can acquire information from the field only by expending a certain amount of energy. Our main result establishes a lower bound on this energy expenditure. 

Our analysis utilizes tools from information theory, quantum field theory in curved spacetimes, general relativity, and thermodynamics to provide insights into relativistic quantum thermodynamics.

The paper is structured as follows: In Sec.\ref{secii}, we provide a detailed review of the communication system employed in this study, focusing on classical channel capacity and energy costs for information transmission in globally hyperbolic spacetimes\cite{Landulfo2016,Landulfo2021}. Section~\ref{seciii} applies Landauer's principle and the second law of thermodynamics to this context, forming the core of our investigation. Finally, Sec.~\ref{seciv} presents our concluding remarks. Throughout, we adopt the metric signature $(-,+,+,+)$ and use natural units where $c=\hbar=1$.

\section{Energy Cost for conveying information}
\label{secii}

In this section, our primary goal is to introduce the system of interest and establish notation. We outline the communication process under consideration and analyse the energies involved, which form the foundation for subsequent discussions in the paper. Specifically, we define the communicating partners and describe the communication channel they use, including its classical capacity $\mathcal{C}$. Furthermore, we discuss the energy changes within the global system (comprising the partners and the channel) during the communication process. This section builds upon the developments presented in Refs.~\cite{Landulfo2016,Landulfo2021}, and interested readers are referred to these works for more detailed explanations. For a deeper understanding of quantum field theories in curved spacetimes, see also Ref.\cite{Waldqft}.

To start, let's consider two communicating partners, Alice and Bob, each possessing a two-level quantum system (a qubit). The communication channel is physically represented by a quantum scalar (massless) field $\phi$. The entire system resides in a globally hyperbolic (asymptotically flat) spacetime $(\mathbb{M}, \mathbf{g})$, where $\mathbb{M}$ denotes the four-dimensional spacetime manifold and $\mathbf{g}$ is a Lorentzian metric.

The communication process unfolds as follows~\cite{Landulfo2016,Landulfo2021}: Alice wishes to transmit information to Bob, encoded in her qubit state $\rho^{A}_{\infty} \in \mathcal{H}_{A}$, which was prepared locally at the infinite past. To encode this information into the communication channel, the field $\phi$, Alice interacts her qubit with the field over a specific time interval $\Delta t_{A}$. This interval is measured relative to a Cauchy hypersurface $\Sigma_t$, where $t$ is a parameter. Once the information is encoded into the field state, the dynamics of the field will mediate its transmission to Bob.

Bob, aiming to retrieve the information sent by Alice, initially prepares his qubit in the quantum state $\rho^B_{-\infty}\in\mathcal{H}_B$ and switches on its interaction with the field for a duration $\Delta t_{B}$, also measured with  respect to a parameter $t$ on some Cauchy hypersurface $\Sigma_t$. It's important to note that Bob's qubit interaction cannot be excessively strong due to potential decoherence effects.

The dynamics of the field $\phi$ is determined by the action
\begin{equation}
    S = -\frac{1}{2}\int_{\mathbb{M}} \dd^4 x \sqrt{-g} \left( \nabla_\mu \phi\nabla^\mu  \phi\right), \label{action}
\end{equation}
where $g \equiv \textrm{det}(g_{\mu\nu})$ stands for the determinant of the metric. From this we obtain the Klein-Gordon equation
\begin{equation}
     \nabla_\mu \nabla^\mu  \phi= 0. \label{kleingordoneq}
\end{equation}
The dynamical evolution of $\phi$ can then be determined by the specification of smooth functions $\phi(t, \textbf{x})$ and $\pi(t, \textbf{x})$ on $\Sigma_t$, where $\textbf{x} \equiv (x^1, x^2, x^3)$. Since we are working in a globally hyperbolic spacetime, this is always possible. 

The canonical momentum $\pi$ is then defined as
\begin{equation}
    \pi = \pdv{S}{\dot{\phi}},
\end{equation}
and the pair $\left(\phi(t,\textbf{x}), \pi(t,\textbf{x})\right)$ represents the state of the field at time $t$, which can be described as a point in the phase space $\mathcal{M}$
\begin{equation}
    \mathcal{M} \equiv \left\{ \phi: \Sigma_t \rightarrow \mathbb{C}, \pi: \Sigma_t \rightarrow \mathbb{C} \,|\, \phi,\pi \in C_0^\infty(\Sigma_t) \right\},
\end{equation}
with $C_0^\infty(\Sigma_t)$ representing the set of infinitely differentiable compact support functions on $\Sigma_t$. 

By following the usual quantization procedure, we promote $\phi$ and $\pi$ to operators satisfying the equal time canonical commutation relations
\begin{equation}
     \left[ \phi(t, \textbf{x}), \phi(t, \textbf{x'}) \right]_{\Sigma_t} =\left[ \pi(t, \textbf{x}), \pi(t, \textbf{x'}) \right]_{\Sigma_t} = 0
\end{equation}
and
\begin{equation}
    \left[ \phi(t, \textbf{x}), \pi(t, \textbf{x'}) \right]_{\Sigma_t} = i \delta^3(\textbf{x}, \textbf{x'}).
\end{equation}
Using these operators we define a symplectic structure $\Omega: \mathcal{S}^{\mathbb{C}} \times \mathcal{S}^{\mathbb{C}} \rightarrow \mathbb{C}$ as
\begin{equation}
    \Omega\left( [\phi_1, \pi_1], [\phi_2, \pi_2] \right) \equiv \int_{\Sigma_t} (\pi_1 \phi_2 - \pi_2 \phi_1) \dd^3\textbf{x},
\end{equation}
where each one of the pairs $(\phi_1, \pi_1)$ and $(\phi_2, \pi_2)$ leads us to a unique element $\psi$ of $\mathcal{S}^{\mathbb{C}}$ ---$\mathcal{S}^{\mathbb{C}}$ is the space of complex solutions to Eq.~\eqref{kleingordoneq}.

The symplectic structure $\Omega(\psi_1, \psi_2)$, with $\psi_1, \psi_2 \in \mathcal{S}^{\mathbb{C}}$, can then be used to define the Klein-Gordon inner product as
\begin{equation}
    \braket{\psi_1}{\psi_2}_{\text{KG}} \equiv - i \Omega(\overline{\psi_1}, \psi_2), \label{kginnerproduct}
\end{equation}
which is not positive definite on $\mathcal{S}^{\mathbb{C}}$. Therefore, the one-particle Hilbert space, $\mathcal{H}$, must be chosen as a subspace of $\mathcal{S}^{\mathbb{C}}$, where the inner product in Eq.~\eqref{kginnerproduct} is positive definite. Additionally, the space of complex solutions is comprised as $\mathcal{S}^{\mathbb{C}} \backsimeq \mathcal{H} \oplus \overline{\mathcal{H}}$, with $\overline{\mathcal{H}}$ being the dual Hilbert space.

Defining now a test function $f \in C^{\infty}_0(\mathbb{M})$, we can describe the generalised solution to Eq.~\eqref{kleingordoneq} as an operator-valued distribution. Formally, this operator is a mapping that associates each test function with an operator. More precise, let us define $\mathcal{S} \in \mathcal{S}^{\mathbb{C}}$ as the space of real solutions, the projection operator $K: \mathcal{S} \rightarrow \mathcal{H}$ and the map $E: C^{\infty}_0(\mathbb{M}) \rightarrow \mathcal{S}$, which acts on the test functions $f$ such that 
\begin{equation}
    Ef(x) \equiv Af(x) - Rf(x),
\end{equation}
where $Af$ and $Rf$ are the advanced and retarded solutions, respectively, to the non-homogeneous field equation and $Ef$ is a solution to Eq.~\eqref{kleingordoneq}. Therefore, the quantum field operator is defined as an operator-valued distribution for some test function $f$ as
\begin{equation}
    \phi(f) \equiv i\left[ a(\overline{KEf}) - a^\dagger(KEf) \right],
\end{equation}
and satisfies the relation
\begin{equation}
    \left[ \phi(f_1), \phi(f_2) \right] = -i\Delta(f_1, f_2) \mathbb{I}, \label{phicommutator}
\end{equation}
with $f_1, f_2 \in C^{\infty}_0(\mathbb{M})$, $\mathbb{I}$ represents the identity operator while $\Delta(f_1, f_2)$ is defined as~\cite{Waldqft}
\begin{equation}
    \Delta(f_1, f_2) \equiv \int_{\mathbb{M}} \sqrt{-g} f_1(x)Ef_2(x).
\end{equation}
The problem with this method is that it involves arbitrarily many choices of Hilbert spaces and, thus, also of vacuum and particles representations. Fortunately, we can circumvent this problem by using an algebraic approach, which is a formulation of the quantum field theory that provides a powerful tool for understanding the dynamics and properties of quantum fields (see Ref.~\cite{Waldqft} for more details). In short, while in the usual approach we define states as vectors in Hilbert spaces and observables as operators acting on these spaces, in the algebraic approach we build operators as elements of an algebraic space over which the states will act by means of the identification of a number to each operator~\cite{Waldqft}. Mathematically, let $\mathcal{W}(\mathbb{M})$ be the exponential version of the algebraic algebra of the fundamental observables. The generators of the elements of this Weyl algebra are defined as 
\begin{equation}
    W(Ef) = e^{i\phi(f)},
\end{equation}
and satisfies the relations $W^*(Ef) = W(-Ef)$, $W[E(\nabla^\mu \nabla_\mu - m^2)f] = \mathbb{I}$ and $W(Ef_1)W(Ef_2) = e^{i\Delta(f_1,f_2)/2}W(Ef_1 + Ef_2)$, with $f_1, f_2 \in C^{\infty}_0(\mathbb{M})$. Additionally, we define the algebraic quasi-free state as a positive and normalised linear functional $\omega: \mathcal{W}(\mathbb{M}) \rightarrow \mathbb{C}$, such that
\begin{equation}
    \omega_\nu[W(Ef)] = e^{-\nu(Ef, Ef)/2},
    \label{eq:alg_state} 
\end{equation}
with $\nu$ being an inner product on $\mathcal{S}$ satisfying the relation
\begin{equation}
    \nu(Ef_1, Ef_1)\nu(Ef_2, Ef_2) \ge \frac{1}{4} |\Omega(Ef_1, Ef_2)|^2.
\end{equation}
In this way, we can describe the states of the communication channel without making any preferred choices of Hilbert spaces~\cite{Waldqft}.

We are interested in the maximum amount of information that Alice can reliably convey to Bob. In other words, we are interested in the capacity of the communication channel. As mentioned before, we need to determine the dynamics of our system. Let us then employ the above formalism in order to determine the time evolution of the state of Bob's qubit, which is the receiver. 

If $\rho_{-\infty}^A$ and $\rho_{-\infty}^B$ are the initial states of Alice and Bob qubits, respectively, we can write the system initial state as $\rho_{-\infty} \equiv \rho_{-\infty}^A \otimes \rho_{\infty}^B \otimes \rho_{\omega}$, where $\rho_\omega$ is the density operator associated with the field algebraic state $\omega_\nu$ (see Eq.~\eqref{eq:alg_state}). Also, we define $\omega_\nu[W(Ef)] \equiv \Tr{\rho_\omega W(Ef)}$. 

Time evolution is governed by the total Hamiltonian of the system, which can be written as
\begin{equation}
    H(t) \equiv H_{\phi}(t) + H_{\text{int}}(t),
\end{equation}
where $H_{\phi}(t)$ is the Hamiltonian of the field and $H_{\text{int}}(t)$ is the Hamiltonian associated with the qubits' interaction with the field. It is mathematically more convenient to change to the interaction-picture with respect to the free Hamiltonian. In this representation, the time evolution operator takes the form
\begin{equation}
    U \equiv \vec{T} \exp{-i\int_{-\infty}^{\infty} \dd t \: H_{I}(t)},
\end{equation}
where $\vec{T}$ is the time ordering operator while $H_{I}(t)$ is the interaction picture representation of the Hamiltonian. 

Under these definitions, the final state of the system is given by $\rho_{+\infty} \equiv U \rho_{-\infty}U^\dagger$, from which we determine Bob's qubit final state by tracing out the field and Alice's qubit degrees of freedom
\begin{equation}
    \rho^B = \Tr_{A,\phi}\left\{U \rho_{-\infty}^A \otimes \rho_{-\infty}^B \otimes \rho_{\omega} U^\dagger\right\},
\end{equation}
whose explicit expression, that is not important four our purposes here, can be found in Ref.~\cite{Landulfo2016}. This is the state where the information transmitted by the channel is codified. So, we can think about this state as a quantum memory.

In our investigation, we focus on examining the balance between the energy expended in the communication process and the energy Bob can generate using the acquired information. Therefore, our first task is to quantify the amount of information available to Bob. In the ideal scenario we are considering, this is represented by the channel capacity $C$. By choosing the initial state for Bob's qubit in such a way that the signalling amplitude of the communication is maximised, and using the Holevo-Schumacher-Westmoreland~\cite{Wilde} theorem, the classical capacity of the quantum channel is given by~\cite{Landulfo2016}
\begin{equation}
    \mathcal{C} = S\left( \frac{1}{2} + \frac{\nu_B}{2}|\cos[2\Delta(f_A, f_B)]| \right) - S\left( \frac{1}{2} + \frac{\nu_B}{2}\right), \label{channelcapacity}
\end{equation}
where $S$ is the Shannon entropy and $\nu_B$ is defined as
\begin{equation}
    \nu_B \equiv \omega_\nu[e^{i\phi(2f_B)}], 
\end{equation}
while the indexes $A$ and $B$ labelling the test functions $f_{A}$ and $f_{B}$ stands for Alice and Bob, respectively.

The channel capacity is the maximum rate at which one can reliably convey information. Therefore, Eq.~\eqref{channelcapacity} represents the maximum amount of classical information Bob can get from Alice per use of the quantum channel. Since we are interested in a lower bound of the energy Bob must spend in order to acquire the information, we assume that this is the case.

Some comments are in order here. First, from Eqs.~\eqref{phicommutator} and~\eqref{channelcapacity} we observe that if Alice and Bob are not causally related, i.e., when the spacetime causality makes it impossible for Alice and Bob to have any influence over each other, $\Delta(f_A, f_B) = 0$ and, consequently, the channel capacity vanishes, as it should. If they are causally related, then $\Delta(f_A, f_B) \neq 0$ and $\mathcal{C} > 0$, such that it will be possible for them to communicate over this channel~\cite{Landulfo2016}. In this last case, Bob's final state will contain the amount of information given by $\mathcal{C}$.

Now that we know how much information is available to Bob, we need to understand the energy balance related to this process. We briefly discuss this now and point the reader to Ref.~~\cite{Landulfo2021} for more details.

We want to study how the energy of the total system (two qubits plus the field) changes in time when the state evolves from $\rho_{-\infty}$ to $\rho_{+\infty}$. In order to do this, we just need to compute the total energy variation of the system, which is simply given by~\cite{Landulfo2021}
\begin{equation}
    \Delta E \equiv \expval{H(+\infty)}_{\rho_{+\infty}} - \expval{H(-\infty)}_{\rho_{-\infty}},
\end{equation}
with $\expval{\cdot}_{\rho}$ representing the expectation value taken with respect to the state $\rho$. 

Note that since the qubits interact with the field for a finite amount of time, the interaction part of the Hamiltonian does not contribute, and the total energy change can be recast into the form
\begin{equation}
    \Delta E = E_\phi + E_A + E_B + E_{AB},
\end{equation}
whose formal expressions can be found in Ref.~\cite{Landulfo2021}. Physically, $E_{\phi}$ is the contribution coming from the effect of particle creation due to the dynamic nature of the metric. $E_A + E_B$ are the energies arising from the work that must be performed in order to turn on and off the interaction between the field and the qubits. This term is a function of the qubit trajectories, the coupling constants and the metric. The last term, $E_{AB}$ is the contribution associated to the communication process itself. This depends on the metric, the relative motion between Alice and Bob and on the initial states of the qubits. Such dependence can be tailored to make $E_{AB} = 0$, while maximising the channel capacity, by a convenient choice of the initial state of Bob's qubit~\cite{Landulfo2021}. Remember that Alice's qubit state cannot be fixed since it contains the information she wants to convey to Bob. Therefore, the total change in energy takes the form
\begin{equation}
    \Delta E = E_\phi + E_A + E_B.
    \label{eq:energy}
\end{equation}

This result is particularly interesting because it tells us that we can convey an arbitrary amount of information without any extra energy cost. This is the main fact on which the results presented in the next section are build. 

\section{Information driven heat engine}
\label{seciii}

Let's consider the simplest scenario of inertial Alice and Bob in Minkowski spacetime, as it is sufficient to illustrate our argument. We assume the initial state of the field is the vacuum, thus avoiding additional noise from finite temperature. In this case, since the metric remains unchanged, $E_\phi$ vanishes, and the only contribution to the energy change comes from the coupling of the qubits with the field. For efficient communication, Alice needs to strongly interact her qubit with the field. This energy is provided by a battery in Alice's lab and is not transmitted to Bob~\cite{Landulfo2021}. Therefore, since our interest lies in what happens at Bob's location, the only relevant energy is $E_B$. Here is our contribution: We apply both the Landauer principle and the second law at Bob's laboratory to show that there must be a bound on Bob's ability to couple his qubit with the field to prevent him from violating the second law.

The argument proceeds as follows. Suppose Bob has a heat engine in his lab, with two finite-size reservoirs at the same temperature. The second law states that it is impossible to extract work from this engine. Now, assume Bob has a memory (in equilibrium with the local environments) where he stores the information received from Alice. Bob can erase this information by allowing the memory to reversibly thermalize with one of the environments, using an arbitrarily small amount of energy. It is important to mention here that Bob has to invest energy in order to acquire information, thus changing the state of his qubit (memory). After this, Bob can simply couple his qubit with his environment without any extra energy environment. The information will be erased and there will be a flux of energy (in the form of heat) from the qubit to the environment. So, there must be a cost of acquiring information. And that cannot be arbitrarily small, but must respect Landauer's bound.

The heat flux generated in this way will increase the energy of the environment. Although the effect is small, it is present. Now, Bob has a heat engine with a temperature gradient, from which work can be extracted. We argue that, since no energy was transmitted along with the information, there must be a lower bound on the energy Bob spends to couple his memory (qubit in our case) with the field to acquire the information. Otherwise, Bob would be able to extract work without investing energy, violating the second law. The aim of this section is to compute this bound.

Bob's qubit must interact with the field weakly and for a short duration to avoid decoherence. This coupling can be adjusted to maximize information transfer without incurring additional energy costs. Recall that in this scenario, information is transferred without energy flowing through the field. Bob can then recover an arbitrary amount of information by expending only a minimal amount of energy.

In the case of the Minkowski spacetime, it is possible to analytically compute the change in the energy of the qubits~\cite{Landulfo2021}
\begin{equation}
    E_X = \frac{\lambda_{X}^{2}}{\tau_{X}},
\end{equation}
where $\lambda_{X}$ is the coupling constant of qubit $X$ with the field while $\tau_{X}$ is the time scale associated with the process of switching on and off this interaction. This is a trade-off relation between energy and time. From here on we assume that $\tau_{A} = \tau_{B} = \tau$ is a fixed time scale that only depends on the switching the detectors. Therefore, the only important variable here is the coupling constant between the qubit and the field.

Now, as mentioned before, we assume that Alice has a local battery that provides her the necessary energy to turn on and off the coupling of her qubit with the field. In order to maximise the communication rate, we should choose $\lambda_{A}$ as big as possible, implying that Alice can transfer the information to the field very efficiently, while spending a reasonable amount of energy. 

The key focus is on Bob in this scenario. We are considering that Bob possesses a local battery that he utilizes to connect his qubit to the field (to acquire information). Additionally, Bob has a heat engine with two finite environments at the same temperature, denoted as $T_{c}$. It's important to note that this system is contained within Bob's laboratory, where both quantum mechanics and thermodynamics are applicable, thus presenting no ambiguity to defining physical quantities like temperature. The second law of thermodynamics states that to effectively operate the heat engine, Bob must establish a temperature gradient. While he could use his own battery to create this gradient and operate the engine to generate work, Bob can also employ the information he got from Alice, converting it into useful work using the heat engine. This is where the Landauer principle comes into play.

Landauer principle states that in order to erasure an amount $\mathcal{I}$ of information we must dissipate energy in the form of heat $Q$ such that
\begin{equation}
Q \geq \beta^{-1} \mathcal{I}\ln 2,
\end{equation}
where $\beta$ is the inverse temperature of the environment where the heat goes~\cite{Landauer}.

Now, by assuming that the communication process is performed in the best possible way, Bob gets the amount of information that equals the channel capacity $\mathcal{I} = \mathcal{C}$. So, by reversibly erasing this information, the equality in Landauer's principle is achieved and Bob can increase the temperature of one of his environments as
\begin{equation}
    T_{c} \rightarrow T_{h} = T_{c} + \frac{Q}{c_{T}} > T_{c},
    \label{eq:gradient}
\end{equation}
Where $c_{T}$ represents the thermal capacity of the environment. This can be achieved, for instance, by weakly coupling the qubit to the environment and allowing the system to reversibly thermalize, a process that can be achieved with an arbitrarily small energy cost. It is important to note that this occurs in Bob's laboratory, thus eliminating any issues related to coupling with gravity. The specific method by which this process occurs is not crucial. The key point is that some heat will be transferred to the environment during the erasure process, raising its temperature and creating the necessary temperature gradient to operate the heat engine. Given that Bob is assumed to achieve the exact channel capacity, this represents the maximum temperature gradient he can create.

The Carnot efficiency of this engine is simply given by
\begin{equation}
    \eta = 1 - \frac{T_{c}}{T_{h}} > 0.
\end{equation}
Using Eq.~\eqref{eq:gradient}, which is a consequence of the Landauer principle, we obtain
\begin{equation}
    \eta = 1 - \left[ 1 + \frac{\mathcal{C}\ln{2}}{c_{T}} \right]^{-1}.
\end{equation}

This implies that the maximum work extracted from the heat engine is given by $W=\eta Q > 0$, which is a function of the relativistic channel capacity and, thus, of the metric and the trajectories followed by Alice and Bob.

Therefore, by erasing the information he received from Alice, Bob can operate an information-fueled heat engine, from which he can extract work $W$, while expending an arbitrarily small amount of energy in the process. However, as discussed in the preceding section, Alice can convey an arbitrary amount of information to Bob with no extra energy cost for him than $E_{B}$. We then conclude that, in order for the second law to be obeyed, we must have
\begin{equation}
    W \leq E_{B},
\end{equation}
since what happens in Alice side does not matter to the heat engine. This directly implies a lower bound on the interaction strength between Bob's qubit and the field
\begin{equation}
    \lambda_{B}^{2} \geq \tau \mathcal{C}T_c\left[1-\frac{1}{1 + \mathcal{C}\ln 2/c_{T}}\right]\ln 2.
    \label{eq:bound}
\end{equation}
Note that such bound depends on the spacetime metric and also on the Alice and Bob trajectories, since these variables determine the channel capacity.

\section{Discussion}
\label{seciv}

In this contribution, we consider a relativistic communication scenario in which Alice sends information to Bob via a massless scalar field. The information is encoded and decoded through local interactions between the field and Bob's and Alice's detectors (qubits). An energy analysis revealed that the energy cost of this protocol is concentrated in the coupling of Alice's and Bob's qubits to the field, with no additional energy required for information transmission~\cite{Landulfo2021}. In this context, we considered that Bob operates a locally reversible heat engine. By applying Landauer's principle and the second law of thermodynamics, we derive a lower bound on the strength of the coupling between Bob's qubit and the field. This bound depends on the channel properties and the spacetime metric.

Some comments on our assumption of a finite-size environment are necessary here. The work extracted from the engine is indeed slightly less than what we previously considered. However, the general arguments remain valid as we are considering the best-case scenario. In this case, any amount of work that Bob can extract from the engine must set a limit on the energy he must spend to couple his qubit with the field. Furthermore, we are considering the case where the heat engine operates reversibly ---zero power output--- which implies that the environments are always in equilibrium, even with temperature variations. Note that our reference temperature in Eq.~\ref{eq:bound} is the cold, initial, one.

The predicted effect is expected to be very small. The environments were considered finite since an infinite environment would have infinite thermal capacity, resulting in no temperature change. However, every physical system is finite, and this assumption is crucial for ensuring locality and consistency in defining thermodynamic laws in curved spacetimes. Landauer's bound predicts that $\beta^{-1}\ln 2$ of heat will be generated per bit erasure, so even with finite environments, a strong effect is not expected. Nevertheless, the important message is that the effect must exist for the second law of thermodynamics to hold in Bob's laboratory.

One consequence of this lower bound is that Bob's qubit will experience unavoidable decoherence. For any finite coupling strength, a finite amount of time is required for information transfer from the field to the qubit. During this time, the qubit interacts with the quantum field, leading to decoherence. This reduces the amount of information Bob receives and, consequently, the extracted work.

To illustrate our argument, we considered the simplest case of two qubits in flat spacetime. However, according to the general theory, if Alice and Bob are not causally disconnected, a positive channel capacity is always achievable, implying that our result holds for general spacetimes. The general case includes energy associated with particle creation from the vacuum due to metric changes, which tends to destroy the information flowing through the channel and decrease the heat engine's efficiency. Despite this, the effect is not expected to vanish unless the channel capacity does.

This work raises several questions, including the study of general spacetimes, especially those with event horizons. Another important issue is the propagation of information through the channel and its relation to energy balance during the dynamics. Additionally, what happens when transmitting quantum information instead of classical information, as considered in this work? These questions will be the subject of future research.

\begin{acknowledgments}
This work was supported by the National Institute for the Science and Technology of Quantum Information (INCT-IQ), Grant No. 465469/2014-0, the National Council for Scientific and Technological Development (CNPq), Grant No. 308065/2022-0, and the Coordination of Superior Level Staff Improvement (CAPES).
\end{acknowledgments}

\appendix



\end{document}